\documentclass[12pt]{article}
\usepackage{amsmath}
\usepackage{graphicx,psfrag,epsf}
\usepackage{enumerate}
\usepackage{natbib}
\usepackage{bbm}
\usepackage{caption}
\usepackage{subcaption}
\usepackage{tikz}
\usepackage{makecell}
\usepackage{amssymb}
\usepackage{multirow}
\usepackage{xcolor}
\usepackage[scr=boondox]{mathalpha}
\usepackage[hidelinks]{hyperref}
\usepackage[linesnumbered,ruled,vlined]{algorithm2e}
\usepackage{algpseudocode}
\usepackage{geometry}
\usepackage{booktabs}

\SetCommentSty{mycommfont}

\SetKwInput{KwInput}{Input}                
\SetKwInput{KwOutput}{Output}              

\newcommand{\blind}{0}

\providecommand{\keywords}[1]
{
  \small	
  \textbf{\textit{Keywords:}} #1
}

\date{}

\begin{document}

\def\spacingset#1{\renewcommand{\baselinestretch}%
{#1}\small\normalsize} \spacingset{1}

\newcommand{\revision}{\color{black}}
\newcommand{\revisionii}{\color{black}}

\if0\blind
{
\title{SynthTree: Co-supervised Local Model Synthesis for Explainable Prediction}
  \author{   Evgenii Kuriabov \\
   Department of Statistics, The Pennsylvania State University\\
     \\
  Jia Li \\
    Department of Statistics, The Pennsylvania State University\\
     \\
 }
  \maketitle
} \fi

\if1\blind
{
  \bigskip
  \bigskip
  \bigskip
  \begin{center}
    {\LARGE\bf Title}
\end{center}
  \medskip
} \fi

\maketitle

\bigskip

\begin{abstract}

Explainable machine learning (XML) has emerged as a major challenge in artificial intelligence (AI). Although black-box models such as Deep Neural Networks and Gradient Boosting often exhibit exceptional predictive accuracy, their lack of interpretability is a notable drawback, particularly in domains requiring transparency and trust. 
This paper tackles this core AI problem by proposing a novel method to enhance explainability with minimal accuracy loss, using a Mixture of Linear Models (MLM) estimated under the co-supervision of black-box models. We have developed novel methods for estimating MLM by leveraging AI techniques. Specifically, we explore two approaches for partitioning the input space: agglomerative clustering and decision trees. The agglomerative clustering approach provides greater flexibility in model construction, while the decision tree approach further enhances explainability, yielding a decision tree model with linear or logistic regression models at its leaf nodes. Comparative analyses with widely-used and state-of-the-art predictive models demonstrate the effectiveness of our proposed methods. Experimental results show that statistical models can significantly enhance the explainability of AI, thereby broadening their potential for real-world applications. Our findings highlight the critical role that statistical methodologies can play in advancing explainable AI.

\end{abstract}

\keywords{Explainable machine learning, Black-box model co-supervision, Decision Trees, Mixture of Linear Models, AI Interpretability
}

\spacingset{1.5}

\section{Introduction}
\label{sec:intro}
The topic of explainable machine learning (XML) has attracted rapidly growing interest in recent years~\citep{adadi2018peeking}.
While black-box models such as {\em Deep Neural Networks} (DNN), {\em Random Forest} (RF)~\citep{breiman2001random}, and {\em Gradient Boosting} (GB)~\citep{friedman2001greedy} often excel in prediction accuracy across numerous applications, their adoption is hindered in fields where the explainability of models is paramount. This is particularly evident in medical contexts, where the inability to comprehend the decision-making process of a model can lead to distrust. Similarly, in many scientific disciplines, the primary objective of prediction is to gain insight into underlying phenomena, making the lack of transparency a significant drawback for any learned model. 

In contrast, statistical models such as {\em Classification and Regression Trees} (CART)~\citep{breiman1984classification} and linear models~\citep{hastie2009elements} offer greater explainability but often at the expense of accuracy. In this paper, we aim to synergize the strengths of both classical statistical models and black-box models. Our goal is to achieve high predictive accuracy while preserving the interpretability of the models. Building upon the concept of co-supervision by black-box models as proposed by~\cite{seo2022mixture}, we extend this framework in two significant ways. First, our approach diverges from the dependency on clustering intermediate outputs from a DNN to construct an interpretable model. Instead, in addition to neural networks, we investigate black-box models that are more familiar to the statistics community, such as RF~\citep{breiman2001random} and GB~\citep{friedman2001greedy}. In addition, we explore the potential of utilizing multiple black-box models for enhanced co-supervision. Secondly, and more crucially, we introduce a novel algorithm for creating a mixture of linear models. This algorithm segments the feature space in a manner akin to CART, thereby strengthening the explainability of the resulting model. 

Our method represents a stride towards reconciling the often conflicting goals of accuracy and transparency in predictive modeling. It also exemplifies the application of statistical methodologies to address significant challenges in artificial intelligence (AI), and conversely, provides a paradigm for improving the construction of statistical models by leveraging AI techniques.

\subsection{Background on Explainable Machine Learning}
\label{sec:back}

Despite extensive efforts to improve the interpretability of black-box models, a definitive understanding of ``interpretability'' remains elusive. The term itself is inherently multifaceted and subjective. In existing literature, numerous interpretations of ``interpretability'' have been proposed, yet a cohesive framework that integrates these various perspectives is still lacking. The review by~\cite{doshi2017towards} thoroughly explores this concept, highlighting its complexity and presenting a variety of philosophical perspectives. \cite{allen2023interpretable} discuss the widespread use of machine learning (ML) in processing large and complex datasets across various fields. Specifically, it focuses on interpretable machine learning (IML), which produces human-understandable insights and facilitates data-driven discoveries. 
Several survey articles reviewing the evolution of artificial intelligence (AI) and ML point out model explainability as a major challenge amidst the growing prevalence of these technologies.
~\cite{9645355} provide a comprehensive survey on Explainable AI (XAI), covering design, development, goals, evaluation methods, and the application of XAI to enhance system security. This article underscores the significance of addressing biases in AI predictions and decisions. Furthermore, it offers insights into the current state-of-the-art, the challenges faced, and future prospects in the field of XAI.

Furthermore, \cite{9007737} review explainable machine learning in natural sciences, focusing on three key elements: transparency, interpretability, and explainability. The paper enhances understanding of XAI's varied applications by surveying recent studies that apply machine learning in natural sciences and integrate explainable machine learning with domain knowledge.
In the specific context of drug discovery, ~\cite{jimenez2020drug} explore the potential of deep learning, emphasizing the importance of model interpretability, and advocate for the development of ``explainable'' deep learning methods. The article not only reviews algorithmic concepts in XAI but also foresees future opportunities and applications in molecular sciences, providing insightful contributions to the advancement of drug discovery technologies.

An interesting study by~\cite{10.1145/3351095.3375624} explores how organizations utilize explainability. It is found that current deployments primarily benefit machine learning engineers for model debugging, leading to a gap between explainability in practice and transparency goals for end users. The paper addresses limitations in current techniques, proposes a framework for clear explainability goals, and concludes with a nuanced discussion on concerns, shedding light on practical applications and challenges in organizational contexts.

\subsection{Related Work}
\label{sec:related}

Interpretability in modeling can be pursued through a variety of strategies. One common approach is the use of inherently interpretable models, such as decision trees or rule-based systems. This method enables data scientists to straightforwardly validate the model's decisions. However, a notable drawback is often a reduction in accuracy. 
Another strategy involves implementing complex models in practical scenarios, while supporting their explainability through two primary categories of methods. The first category aims to simplify complex models, yet these models typically do not reach a simplicity level that permits direct explanation. For example, DNN with stimulated learning (\cite{7404853}) injects information into each of the hidden layers with two primary objectives: facilitating post-training visualization of these hidden layers and establishing a degree of certainty on the specialized recognition functions of specific regions within the hidden layer. However, as the end-product it is still a complicated structure of a DNN even if the hidden layers are better explained. With the growth of a depth and large size of each layer the interpretability will decrease. Another approach was proposed by \cite{schneider2021explaining}, which explains neural networks by decoding layer activations. They initially determined directions using a generator and a ``layer selectivity'' heuristic. Subsequently, they formulated post-hoc explanations for the encoded content. One challenge for this approach is that the internal layers of a network may not accommodate useful concepts while 
single directions within the activation space may correspond to unrelated concepts. The second category focuses on interpreting complex models in more confined contexts, for instance, offering ``local explanations'' for specific instances or samples~\citep{ribeiro2016model} and elaborating on the interactions between features. In contrast to these approaches, our methodology is designed to provide a ``global explanation.'' We achieve this by constructing an interpretable model that approximates the functionality of its more complex counterparts.

Some approaches address interpretability during the model’s training phase. These approaches first define what constitutes effective interpretability. For example, attention-based methods (\cite{bahdanau2014neural,vaswani2017attention}) introduce attention scores, and feature selection methods implement specific error functions or penalties, such as the $L_0$ penalization suggested by~\cite{tsang2018neural}. In model-agnostic approaches, interpretability is attained post-hoc by analyzing a fully trained model using interpretable measures, as seen in methods like local linear surrogate models~\citep{ribeiro2016should} and Bayesian non-parametric mixtures~\citep{guo2018explaining}, which pinpoint crucial features for specific instances.

How interpretation of models is provided varies over a wide spectrum, ranging from visualizing results to generating human-readable rules, feature selection, and constructing prototype cases. ~\cite{molnar2020interpretable} discusses the pros and cons of these various methods. The selection of an interpretation method often depends on the application field, with different methods holding varying importance in different areas, e.g., image processing and analysis of tabular data. 

Our methodology aligns closely with the works of ~\cite{seo2022mixture} and ~\cite{kunzel2022linear}. Building upon the foundation laid by~\cite{seo2022mixture}, our approach introduces substantial modifications and enhancements. The model we have developed bears a resemblance to the {\em Linear Regression Trees} (LRT) developed by~\cite{kunzel2022linear}. 
Both methodologies produce a decision tree, with each leaf containing a linear model. Despite this similarity, our method of constructing the decision tree is markedly different from that of ~\cite{kunzel2022linear}.
Our approach is more focused on explainability, whereas LRT is more akin to RF. Given the highly competitive accuracy achievable by LRT, we utilize it as one of the black-box models for co-supervision. Additionally, the complexity level of LRT can be adjusted to produce explainable models. We conduct experiments to compare our method with this usage case of LRT~\citep{kunzel2022linear} as well as the approach of ~\cite{seo2022mixture}.

The rest of the paper is structured as follows. We provide an overview of our approaches and introduce basic definitions and notations in Section~\ref{sec:pre}. The SynthTree algorithm is presented in Section~\ref{sec:alg}. Experimental results are provided in Section~\ref{sec:exp}. Finally, we conclude and discuss future work in Section~\ref{sec:conclude}.

\section{Preliminaries}
\label{sec:pre}

Our main idea is to predict the response by a {\em Mixture of Linear Models} (MLM) and estimate MLM via co-supervision from pre-trained black-box models. Our method involves partitioning the input space into distinct regions and fitting a linear model to each. For binary classification, logistic regression is used. While a linear model may not always be immediately interpretable, their inherent explainability is widely accepted.  This explainability can be further enhanced by imposing LASSO-type sparsity penalties~\cite{https://doi.org/10.1111/j.2517-6161.1996.tb02080.x}. Additionally, the robustness of these models can be validated via established hypothesis testing methods. The division of the input space into regions can be conducted as either a hard or soft separation, with the latter assigning weights to each region at individual data points. When these regions are straightforward to explain, the overall interpretability of the MLM is high. The concept of co-supervision with black-box models, as investigated by~\cite{seo2022mixture}, demonstrates that this approach effectively narrows the accuracy gap typically observed with black-box models. In essence, the use of MLM serves as a foundation for achieving explainability, while co-supervision with black-box models ensures minimal compromise in prediction accuracy.

The effectiveness of MLM in terms of explainability hinges significantly on the clarity and interpretability of the regions into which the input space is divided. The method employed by~\cite{seo2022mixture} involves a bottom-up process, aggregating smaller, quantized sections of the input space into larger areas. These aggregated areas are referred to as {\em EPICs} ({\em Explainable Prediction-induced Input Clusters}). However, one notable limitation of this approach is the potential lack of explainability in the resulting regions. Specifically, the process of merging smaller areas can lead to the formation of fragmented and complex regions that are challenging to interpret and represent. To address these challenges, subsequent methods such as dimension reduction, visualization techniques, and the identification of simpler, more describable sub-regions have been considered.

\subsection{Overview of Approaches}
In this paper, we aim to partition feature space into regions that are straightforwardly explainable, setting our work apart from the approach of ~\cite {seo2022mixture}. Our new algorithm exploits a tree structure for partitioning the space, formed under a principle akin to that of CART~\citep{breiman1984classification}. Each region in our model corresponds to a leaf node of the tree and is designed to be easily interpretable. The final model is a decision tree with each leaf node containing a linear model. Theoretically, this model can be regarded as an MLM, wherein the mixture weights at any given point are either one or zero, contingent upon the leaf node that the point belongs to.  A key idea in the construction of MLM is to partition the input space based on the similarity of local models estimated for the neighborhoods around each data point. Points assigned to the same region will be used to estimate a single linear model. As a result, the construction of MLM is technically a process to synthesize local models grouped in the same region although the direct goal is not model synthesis. We thus call this new algorithm {\em Decision Tree by Co-supervised Local Model Synthesis} and, in short, {\em SynthTree}. 

We also investigate a relatively straightforward extension of the approach by~\cite {seo2022mixture}, in particular, to apply co-supervision in a more general setting. Instead of relying on DNN and cascaded clustering of the outputs of intermediate layers of DNN, we explore co-supervision by RF and GB combined with k-means clustering in the original input space.  Since we employ the same method to generate the regions, we also call them EPIC. Although this aspect of our research does not represent a major methodological advance, it makes several significant contributions. Primarily, it assesses the applicability of co-supervision in contexts where black-box models, other than DNNs, are prevalent. This exploration could broaden the scope of co-supervision as a versatile training strategy. Furthermore, we investigate whether the integration of multiple black-box models in the co-supervision process can enhance the predictive accuracy of MLM. In line with the terminology of~\cite{seo2022mixture}, we refer to this method as {\em MLM-EPIC}.

\subsection{Notations and Definitions}
Denote the independent variables (or covariates) by $X\in\mathbb{R}^{p}$ and the dependent variable by $Y\in\mathbb{R}$. For classification, $Y$ is categorical. Denote the sample space of $X$ by $\mathcal{X}$.
Let $\{y_i\}_{i=1}^n$ and $\{\mathbf{x}_i\}_{i=1}^n = \{(x_{i,1},\cdots,x_{i,p})^T\}_{i=1}^{n}$ be the $n$ observations of $Y$ and $X$. Denote the input data matrix by $\mathbf{X}\in\mathbb{R}^{n\times p}$. For regression, we aim at estimating the following regression function for any $\mathbf{x}\in \mathcal{X}$ (for classification, we  substitute $Y$ by $g(Y)$ via a link function $g(\cdot)$).
\begin{eqnarray}
m(\mathbf{x}) = \mathbb{E}(Y|X=\mathbf{x}) \; .
\label{eq:regress}
\end{eqnarray}
For linear regression, we assume $m(\mathbf{x})=\alpha+\mathbf{x}^T\mathbf{\beta}$. 
Denote the prediction function of a black-box model by $\hat{m}(\bold{x})$ and a piecewise linear model, in particular, MLM, by $\tilde{m}(\bold{x})$. 

For brevity, we discuss in the context of regression, with the understanding that adaptation to classification scenarios is straightforward. Consider the input sample space, denoted by $\mathcal{X}$, is divided into $K$ mutually exclusive and collectively exhaustive sets, $\{\mathcal{P}_1,...,\mathcal{P}_{K}\}$. This segmentation forms a partition of $\mathcal{X}$. Denote the partition by $\mathscr{P}=\{\mathcal{P}_1,...,\mathcal{P}_{K}\}$. The partition $\mathscr{P}$ induces a row-wise separation of the input data matrix $\mathbf{X}$ into $K$ sub-matrices: $\mathbf{X}^{(1)}$, ...,  $\mathbf{X}^{(K)}$.
Within each $\mathcal{P}_k$, $\widetilde{m}(\mathbf{x})$ is approximated by $m_k(\mathbf{x})$, which is referred to as the {\it local linear model}. 

Denote by $I_{\mathcal{P}}(\mathbf{x})$ the indicator function that equals $1$ when $\mathbf{x}\in\mathcal{P}$ and zero otherwise.
Given the partition $\mathscr{P}$, $\hat{m}(\mathbf{x})$ is
\begin{eqnarray}
\hat{m}(\mathbf{x})&=&
 I_{\mathcal{P}_{1}}(\mathbf{x})m_1(\mathbf{x})+\cdots +I_{\mathcal{P}_{K}}(\mathbf{x})m_{K}(\mathbf{x}),\nonumber\\
m_k(\mathbf{x}) &=& \alpha_k + \mathbf{x}^T\mathbf{\beta}_k, \qquad k=1,...,K.
\label{eq:mhat}
\end{eqnarray}

In the estimation of an MLM, two primary challenges arise. The first challenge involves determining the optimal method for partitioning the input space $\mathcal{X}$. The second challenge is that within any given region, the data points are often too sparse to enable a robust estimation of the model $m_k(\mathbf{x})$. To tackle the first challenge, ~\cite{seo2022mixture} suggested forming regions based on the similarity of the local model estimated for the vicinity of each data point, and they discussed the merits of this partitioning criterion. Nonetheless, these neighborhoods are finer in granularity compared to EPICs, exacerbating the second challenge of sparse data for local model estimation. To mitigate this issue, co-supervision from a pre-trained black-box model, particularly a DNN, has been employed. However, in our research, we explore the use of black-box models other than DNNs. Essentially, a highly accurate black-box model is utilized as a surrogate for the true distribution. We leverage this model to create simulated data that supplement the original dataset. The additional data enable the estimation of local models at the fine granularity of neighborhoods surrounding each data point.

\begin{figure}[!htb]
  \centering
  \includegraphics[width=1\linewidth]{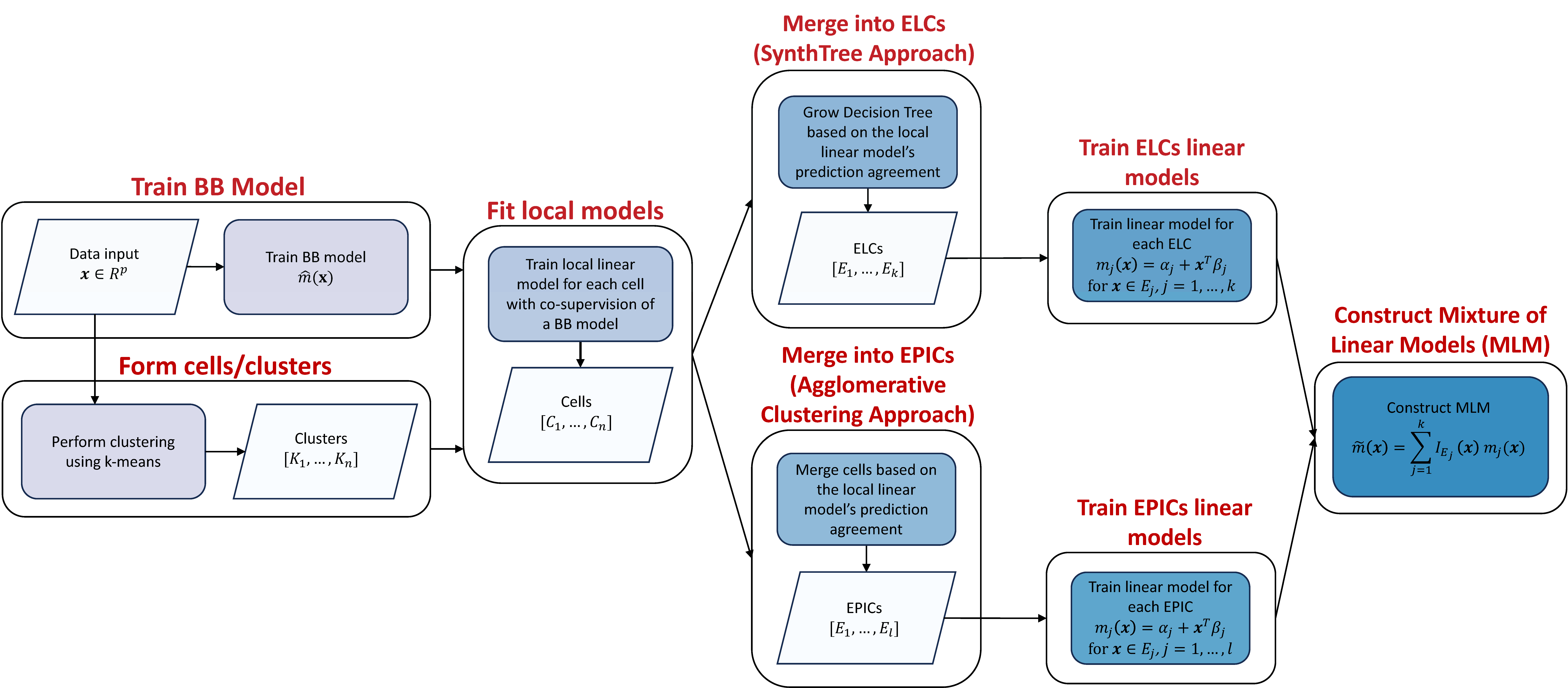}
  \caption{A schematic plot showing the steps of constructing MLM using two approaches developed in this work: SynthTree and MLM-EPIC.}
  \label{fig:figure1}
\end{figure}

\autoref{fig:figure1} provides a schematic illustration of the key steps involved in generating MLM. The creation of partition $\mathscr{P}$ involves a two-step process. Initially, clustering is performed based on the original input, providing preliminary data point clusters and reducing computation in subsequent stages. Also at this stage we train a black-box model, which will be used for co-supervision when constructing linear models for each cluster. This procedure provides us with clusters, referred as {\em cells}, crucial for the next step. Next, we form regions, aka, EPICs, by two optional approaches: tree splitting (SynthTree) or agglomerative clustering (MLM-EPIC). 

Our technical focus is on SynthTree. Each leaf node in the tree corresponds to one region (the equivalence of EPIC) and is referred to as {\em Explainable Leaf Clusters} (ELC). A linear model is then fitted for each ELC/EPIC, using data points and simulated data generated from the black-box model. The interpretability of MLM relies both on the local linear models $m_k(\bold{x})$ and the characterization of the ELCs/EPICs. Shallow trees, in particular, provide simpler explanation.

\section{Algorithm}
\label{sec:alg}


\subsection{Co-supervision by Black-box Models}
\label{sec:cosup}
We train black-box models and apply k-means to cluster data into cells. Denote the cells by $\mathcal{C}_j$, $j=1, ..., J$. Denote the black-box models by $\mathcal{M}_{l}$, $l=1, ..., L$. In our experiments, the black-box models include RF, GB, {\em Multi-Layer Perceptron} (MLP) neural network, and {\em Linear Random Forest} (LRF)~\citep{kunzel2022linear}. LRF and LRT, both included in the {\em Forestry}\footnote{\url{https://github.com/forestry-labs/Rforestry}} software package as noted by~\cite{kunzel2022linear}, differ primarily in how many trees are used in a model: LRT uses a single tree, while LRF employs a typically large collection of trees. Consequently, we categorize LRT as an explainable model and LRF as a black-box model. 

If co-supervision is performed by multiple black-box models simultaneously, we find for each cell $\mathcal{C}_j$, the model $\mathcal{M}_l$ that yields the highest accuracy on the points in this cell. If there is a tie between the models, we choose the one with the highest accuracy over the entire training dataset. With a slight abuse of notation, we denote the chosen index of model $l$ for cell $\mathcal{C}_j$ by $l^{*}(j)$.

Next, we simulate data to augment each cell. Our method of data augmentation is adopted from~\cite{seo2022mixture}. Specifically, let the mean (aka, centroid point) of cell $\mathcal{C}_j$ be $\bar{\mathbf{x}}_j$. Then data are simulated by sampling from the Gaussian distribution with mean $\bar{\mathbf{x}}_j$ and covariance set to a diagonal matrix containing the sample variances of each dimension based on points in this cell. The number of simulated data per cell depends largely on the computational resource. We set it to $100$ in the experiments. For every sample point $\mathbf{x}$, we generate the corresponding response variable by the prediction of black-box model $\mathcal{M}_{l^*(j)}$ chosen for cell $\mathcal{C}_j$.

Based on augmented data in each cell $\mathcal{C}_j$, we apply Lasso to estimate a local linear model, denoted by $g_j(\mathbf{x})$. The local linear models are integrated to create an initial mixture of linear models denoted by $\tilde{g}(\mathbf{x})$:
\[
\tilde{g}(\mathbf{x}) = \sum_{j=1}^{J} I_{\mathcal{C}_j} g_j(\mathbf{x}),
\] where $I_{\mathcal{C}_j}$ is an indicator function that equals $1$ when $\mathbf{x} \in \mathcal{C}_j$ and zero otherwise.


To determine the optimal number of cells $J$, we conduct experiments using various values of $J$ ranging from tens to hundreds, contingent upon the data size. The selection of $J$ is guided by the performance of the final model, specifically choosing the value that yields the highest accuracy within the training dataset.




\subsection{Construct SynthTree}
\label{sec:synth}

A key distinction in the construction of a SynthTree, as compared to a conventional decision tree, lies in the partitioning process. In SynthTree, the partitioning defined by the tree structure is applied to a collection of cell-wise prediction models, denoted as 
$g_j(\mathbf{x})$, $j=1, ..., J$, rather than directly to the sample points. However, it is important to note that the input feature space remains the domain where this partitioning takes place. The process of constructing SynthTree encompasses three main steps:
\begin{enumerate}
\item
    Determining the rule for splitting each node.
    \item
Deciding whether to classify a node as terminal (aka, leaf) or to further divide it.
\item
Assigning a prediction function to each of the leaf nodes.
\end{enumerate}

In the first step, similar to the approach used in CART, we limit our choice of splitting rules to single-variable thresholding at various cutoff values. This is represented as $X_j\leq r_{j,\tau}$, $j=1, ..., p$, $\tau=1, ..., \zeta_{j}$, where $\zeta_{j}$ is the number of cutoff points for the $j$th dimension. To select an appropriate splitting rule, we apply the {\em ``goodness of split''} criterion as used in CART. We then proceed to define this goodness of split for each node, applicable to any chosen splitting rule.

\begin{figure}[!htb]
  \centering
  \includegraphics[width=0.9\linewidth]{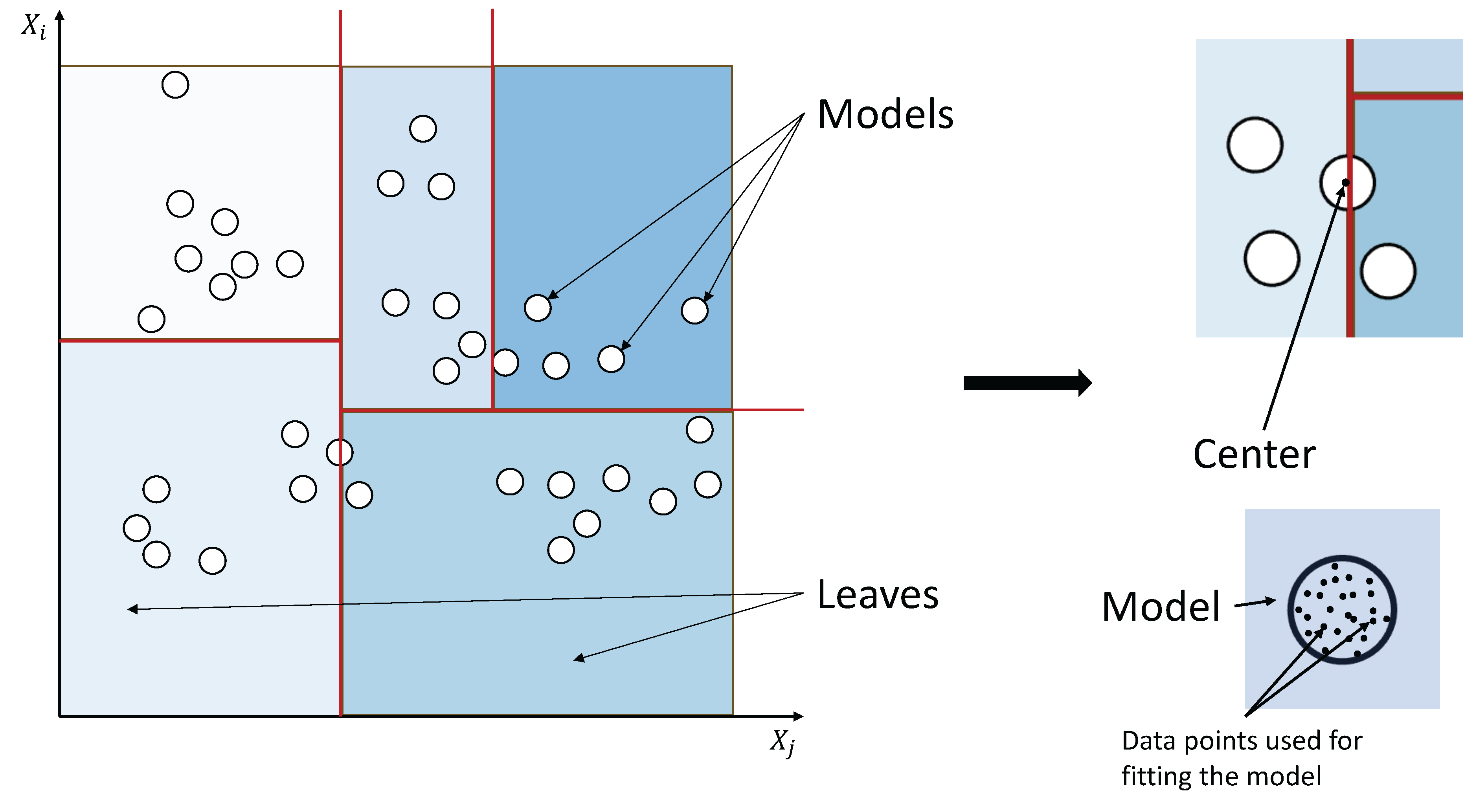}
  \caption{A schematic plot to illustrate the partition of the feature space based on SynthTree.}
  \label{fig:figure2}
\end{figure}

For every cell-wise linear model $g_j$, $j=1, ..., J$, it is estimated from a set of original and augmented data points. Recall the mean of these data points is denoted by $\bar{\mathbf{x}}_j$. To determine whether $g_j$ belongs to a node $t$ in a tree, we check whether $\bar{\mathbf{x}}_j$ belongs to node $t$. It is possible that points used to estimate $g_j$ fall into other nodes. This characteristic of SynthTree is illustrated in Figure~\ref{fig:figure2}. Next, without loss of generality, consider a node $t$ that contains linear models $g_j(\mathbf{x})$, $j=1, ..., J_t$. Following the practice in~\citep{seo2022mixture}, we define a distance between two models, say $g_1$ and $g_2$, as the {\em mutual prediction disparity} in the case of regression and the {\it inverse of $F_1$-score}~\citep{van1979information} in the case of classification. 

Suppose $g_1$ is estimated from data points $\mathbf{x}_{i}$, $i=1, ..., n_1$ and $g_2$ from $\mathbf{x}_{i}$, $i=n_1+1, ..., n_1+n_2$. Then the mutual prediction disparity, $d(g_1, g_2)$ is
\begin{eqnarray*}
d(g_1, g_2)=\frac{1}{n_{1}+n_{2}}\sum_{i=1}^{n_{1}+n_2}(g_{1}(\mathbf{x}_{i})-g_{2}(\mathbf{x}_{i}))^2 \; .
\end{eqnarray*}
For classification, $d(g_1, g_2)$ is defined as
\[ d(g_1,g_2) = \frac{f_p + f_n}{2\cdot t_p} \, , \]
where
\begin{eqnarray*}
t_p &=& |\{\textbf{x}_{i}|g_1(\textbf{x}_{i}) = 1 \; \mbox{and} \; g_2(\textbf{x}_{i}) = 1, \, i = 1, ..., n_1+n_2\}|  \; , \\
f_p &=& |\{\textbf{x}_{i}|g_1(\textbf{x}_{i}) = 1 \; \mbox{and} \; g_2(\textbf{x}_{i}) = 0, \, i = 1, ..., n_1+n_2\}|  \; , \\
f_n &=& |\{\textbf{x}_{i}|g_1(\textbf{x}_{i}) = 0 \; \mbox{and} \; g_2(\textbf{x}_{i}) = 1, \, i = 1, ..., n_1+n_2\}|   \; ,
\end{eqnarray*}
and $| \cdot |$ denotes the cardinality of a set.

Finally, we define the {\em impurity function} of node $t$, denoted as $D(t)$, by the average pairwise distance between $g_j$'s, $j=1, ..., J_t$. A smaller $D(t)$, as a heuristic for defining the impurity function, indicates lower impurity due to greater similarity among models within the same node. Suppose $t$ is split into two child nodes denoted as $t_L$ and $t_R$. Let the proportion of models assigned to $t_L$ be $p_L$ and that assigned to $t_R$ be $p_R$. Both proportions are with respect to the number of models in the parent node $t$. Then, the {\em goodness of split} is computed by
\[
D(t)-p_L\cdot D(t_L)-p_R\cdot D(t_R) \; .
\]
Note that the way we define goodness of fit follows precisely what is used in CART. It is observed that the goodness of split is not always positive. We stop splitting a node if no split can yield a positive goodness of split. In addition, apparently, if the node contains only one model, it cannot be further split. 

Similar to the approach used in standard decision tree construction, SynthTree is developed in a step-wise, greedy fashion, with the aim to maximize the goodness of split at each node. This involves evaluating every potential split from a set of candidates, and selecting the split that offers the highest value of goodness of split. As the growth of the decision tree is not conducted on the original data points but rather on the local linear models, the number of which is usually much smaller than the data size, we expand the tree to its maximum depth. Utilizing models for tree construction provides a notable advantage in terms of computational efficiency.  


In the third step of the tree construction process, we employ Lasso regression to fit a linear model in each leaf node. In particular, we use both the original data and the augmented data, which are identical to those used for training the cell-wise local models (see Section~\ref{sec:cosup}). We pass the data points through the tree, identifying those that land in each leaf. These identified points are then used to fit the linear model in the respective leaf nodes.

\subsection{Right-sized Tree}
\label{sec:prune}

To determine the optimal depth for our decision tree, we investigate two approaches. 
We first grow the deepest possible tree. Suppose the maximum depth of the tree is $\gamma$. In the first approach, named {\it leveled trimming (L-trim)}, we examine subtrees with depth ranging from $1$ to $\gamma$. At each depth in this range, we compute the $10$-fold cross-validation (CV) accuracy based on the training data. The depth yielding the best CV accuracy is selected. 

In the second approach, we adopt the pruning process in CART based on a cost-complexity measure defined in our context. We refere to this method as {\it cost-complexity pruning (CC-prune)}. Consider a tree denoted by $T$ and let its collection of leaf nodes be $\tilde{T}$. The number of leaf nodes is $|\tilde{T}|$. For each node $t$, regardless of whether it is a terminal node, we fit a linear model and compute the total squared error (or error rate for classification) based on training data points assigned to this node, which is called resubstitution error $R(t)$. Denote the linear model fitted for node $t$ by $m_t(\mathbf{x})$. In the case of regression, 
\[
R(t)=\sum_{(\mathbf{x}_i,y_i)\in t}|y_i-m_t(\mathbf{x}_i)|^2 \; .
\]
For classification, let $I(\cdot)$ be the indicator function. Then 
\[
R(t)=\sum_{(\mathbf{x}_i,y_i)\in t}I(y_i\neq m_t(\mathbf{x}_i)) \; .
\]

Let $R(T)$ be the re-substitution error for tree $T$. Then 
\[
R(T)=\sum_{t\in \tilde{T}}R(t)
\]
We denote a branch growing out of $t$ by $T_t$, which includes $t$ and all its descendent nodes. Then the re-substitution error for the branch $T_t$ is
\[
R(T_t)=\sum_{t\in \tilde{T}_t}R(t) \; .
\]

Given a complexity parameter $\alpha$, the {\em cost-complexity measure} is defined as follows:
\begin{enumerate}
    \item 
    For a node $t$, $R_{\alpha}(t)=R(t)+\alpha$.
     \item 
    For a branch $T_t$, $R_{\alpha}(T_t)=R(T_t)+\alpha|\tilde{T}_t|$.
    \item
For a tree $T$,
$R_{\alpha}(T)=R(T)+\alpha |\tilde{T}|$.
\end{enumerate}

For every given $\alpha$, we want to find the subtree $T(\alpha)$ that minimizes $R_{\alpha}(T)$, that is,
\[
R_{\alpha}(T(\alpha))=\min_{T'\preceq T}R_{\alpha}(T') \; .
\]
We indicate a subtree of $T$ using notation ``$\preceq$''. A larger value of $\alpha$ in general results in a smaller subtree $T(\alpha)$. We employ the weakest-link cutting procedure in CART to perform pruning. Start with a tree $T$. Set $T_1=T$ and $\alpha_1=0$ For every node $t\in T_1$, we compute the following function
\begin{eqnarray*}
h_1(t)=\left\{
\begin{array}{ll}
\frac{R(t)-R(T_t)}{|\tilde{T}_t|-1}\, , & t\notin \tilde{T}_1\\
+\infty \, , & t\in \tilde{T}_1
\end{array}
\right .
\end{eqnarray*}

Define the weakest link  $\bar{t}_1$ in $T_1$ as the node such that \[h_1(\bar{t}_1)=\min_{t\in T_1}h_1(t)\] and put $\alpha_2=h_1(\bar{t}_1)$.

When $\alpha$ increases,  $\bar{t}_1$ is the first node that becomes more preferable than the branch $T_{\bar{t}_1}$ descended from it. In fact, $\alpha_2$ is the first value after $\alpha_1 = 0$ that yields a strict subtree of $T_1$ with a smaller cost-complexity measure. We let $T_2=T_1-T_{\bar{t}_1}$, that is, $T_2$ is the subtree of $T_1$ with branch $T_{\bar{t}_1}$ pruned. Once $T_2$ is obtained, we repeat the above process, treating $T_2$ in the same way as $T_1$. The corresponding function $h_2(t)$ may differ from $h_1(t)$ at some nodes because after pruning, the collection of leaf nodes of a branch out of node $t$ may have changed, which affects both $R(T_t)$ and $|\tilde{T}_t|$. In particular, any ancestor node of a pruned branch will in general have an altered value of $h_2(t)$. This pruning process will be repeated until only the root node of the tree is left. 

The original CART decision tree has the theoretical property that the critical values of the complexity parameter $\alpha_k$, $k=1, 2,, ....$, are increasing and the optimal subtrees $T(\alpha)$ are nested when $\alpha$ increases~\citep{breiman1984classification}. In our case, however, as we use a linear model in each leaf node, due to nuances in fitting these linear models, we cannot ensure that $\alpha_k$'s are increasing with $k$ or even being non-negative although that is usually the pattern. We remedy this issue by directly imposing an increasing sequence of $\alpha_k$. Specifically, suppose we have pruned the tree at $\alpha_k$ and proceed to tree $T_k$. In the next round of pruning, we first find the value $\tilde{\alpha}=\min_{t\in T_k, h_k(t)\geq \alpha_k }h_k(t)$, that is, the smallest $h_k(t)$ that is no smaller than $\alpha_k$. We have at least one $\bar{t}_k$ such that $h_k(\bar{t}_k)\leq \tilde{\alpha}$, but there can be multiple $\bar{t}_k$'s with $h_k(\bar{t}_k)\leq \tilde{\alpha}$. The branches at all these nodes will be pruned off. It is assumed that for $\alpha_k\leq \alpha<\alpha_{k+1}$, the optimal subtree $T(\alpha)$ is the same as $T(\alpha_k)$. 

We then use the same cross-validation procedure in CART to select the optimal complexity parameter, denoted as $\alpha^*$. In summary, we start by pruning both the tree constructed from the entire training dataset, denoted by $T^{(all)}$, and the trees generated from each fold in the cross-validation (CV), denoted as $T^{(l)}$, where $l$ represents the specific fold. In our experiments, $10$-fold CV is used. Based on the pruning of $T^{(all)}$, we derive a sequence of candidate values for $\alpha_k$.  For each value of $\alpha_k$, we identify the corresponding optimal tree in the $l$th fold and calculate its error rate on the held-out data in that fold. The optimal $\alpha^*$ is then chosen based on the lowest CV error rate. For a more detailed explanation of this process, we refer readers to~\citep{breiman1984classification}.

We have observed empirically that CC-pruning improves classification accuracy for SynthTree but yields worse results than L-trim in the case of regression. In the next section on experiments, we report results by CC-pruning for classification and those by L-trim for regression. The preference for L-trim or CC-pruning is consistent between accuracy levels achieved on training and test data respectively. As a result, to select between the two schemes, we can compare them based on training accuracy.

\section{Experiments}
\label{sec:exp}





To evaluate prediction results, we use either {\em Root Mean Squared Error} ($RMSE$) or {\em Area Under the ROC Curve} ($AUC$) for regression and classification respectively. Suppose the true and predicted response is $y_i$ and $\hat{y}_i$ respectively, $i=1, ..., n$ ($n$ is the data size). $RMSE$ is defined as
$RMSE = \sqrt{\sum_{i=1}^{n}(y_i - \hat{y}_i)^2/n}$.
A lower $RMSE$ value signifies greater accuracy in prediction. In the context of classification, the ROC curve is a graphical representation of the True Positive Rate (TPR) against the False Positive Rate (FPR) across various decision thresholds. The $AUC$ is the integral of the ROC curve within the range [0, 1]. The AUC, ranging from zero to one, serves as an indicator of classification performance, with higher values denoting superior performance. For each method evaluated, reported results are the mean outcome derived from five distinct random splits of the dataset into training and test sets.

We compared with four black-box models: RF, GB, LRF, and Multi-Layer Perceptron (MLP) neural network. All models were trained using default settings. In particular, for RF and LRF, $100$ and $500$ trees are used respectively. Our methods, SynthTree and MLM-EPIC, were evaluated in two scenarios: using a single black-box model, and using all four black-box models for co-supervision. For clarity in reporting results, we use the following naming conventions. When all black-box models are employed simultaneously for co-supervision, our methods are referred to as SynthTree-INT and MLM-EPIC-INT, respectively. In cases where co-supervision involves a single black-box model, for example, RF, we denote it as SynthTree-RF or MLM-EPIC-RF. For MLM-EPIC, cross validation is used to choose the number of EPICs. Additionally, we tested three explainable methods: CART, LR, and LRT.
See Section~\ref{sec:intro} for more description of these methods.

Eight datasets used in our experiments (four for regression and four for classification) are listed below. For all the datasets, we use dummy encoding for nominal variables. Basic information about these datasets including their data size and dimensions is provided in Table~\ref{tab:table}.

\begin{itemize}
    \item{\textbf{The Cancer Genome Atlas Skin Cutaneous Melanoma (TCGA-SKCM)}: This is a clinical dataset collected to investigate the correlation between cancer phenotypes and genotypes. The target variable in this dataset is the overall survival (OS) status, dichotomized into binary values, with 1 indicating living and 0 deceased.}
    
    \item \textbf{Bikesharing}: 
    This dataset comprises hourly time series information on bike rentals within the Capital Bikeshare system spanning the years 2011 to 2012. The target variable is the total count of rented bikes.    
    
    \item \textbf{Compas}: 
    The COMPAS (Correctional Offender Management Profiling for Alternative Sanctions) dataset contains variables employed by the COMPAS algorithm to assess defendants, alongside their outcomes within a 2-year time frame following the decision. The dataset includes information on over 10,000 criminal defendants in Broward County, Florida. The target variable is ``recidivism occurrence'', which equals 1 for cases involving recidivism and 0 otherwise.
    
    \item \textbf{Abalone}:
    This dataset focuses on predicting the age of abalones based on various physical measurements. The age of abalone is determined by cutting the shell through the cone, staining it, and counting the number of rings through a microscope. Other measurements, which are easier to obtain, are used to predict the age. 
        
    \item \textbf{Road Safety}:
    Data reported to the police about the circumstances of personal injury road accidents in Great Britain from 1979, and the maker and model information of vehicles involved in the respective accident. This version includes data up to 2015. The target variable is the gender of the driver.

    \item \textbf{Upselling}:
    A TunedIT's ARFF-formatted dataset, part of the KDD Cup 2009, which focuses on Customer Relationship Management (CRM) within the marketing realm. Derived from Orange (a French telecom company), this dataset enables the prediction of customer behavior, including the likelihood of churn, appetency for new products, and the potential for up-selling strategies. The response variable is the binary class Up-selling, which involves encouraging customers to purchase more lucrative items or upgrades for a more profitable sale.

    \item \textbf{Servo}:
    A Servo prediction model, also known as a servo control or servo system, is a control system that uses feedback to accurately position or control the motion of a mechanical device, such as a motor or an actuator. The goal of a servo system is to maintain a desired position or trajectory by continuously monitoring the actual position and making adjustments as needed. The response variable, rise time, is the time required for the system to respond to a step change at a set position.

    \item \textbf{California Housing}:
    This dataset contains median house values in California districts from the 1990 U.S. Census. Each instance corresponds to a geographical block group. The original covariates include 8 features: latitude, longitude, median income, house age, average number of rooms, bedrooms, block population, and average house occupancy. We add a new covariate---the average number of non-bedroom rooms---from the given average number of rooms and bedrooms.

\end{itemize}

\begin{table}[!htb]
\centering
\begin{tabular}{c|c|c|c}
\toprule
\multirow{2}{*}
  {Data} & No.  & No. Orig. & No. Trans.  \\
   & Samples & Variables & Variables\\
  \hline
  SKCM & 388 & 34 & 73 \\
  \hline
  Bike Sharing & 17379 & 12 & 16 \\
  \hline
  Compas & 16644 & 17 & 17 \\
  \hline
  Abalone & 4177 & 10 & 10 \\
  \hline
  Road Safety & 111762 & 32 & 32 \\
  \hline
  Upselling & 5032 & 45 & 45 \\
  \hline
  Servo & 167 & 4  & 19 \\
  \hline
  Cal Housing & 20640 & 8 & 8 \\
\bottomrule
\end{tabular}
\caption{Basic information on the eight datasets in the experiments. For each dataset, the data size, number of original predictor variables, and the number of transformed predictor variables are listed (e.g., coding of nominal variables).}
\label{tab:table}
\end{table}

\begin{table}[!htb]
\centering
\begin{tabular}{l|cc|cc|cc|cc}
\toprule
\multirow{2}{*}{Model} & \multicolumn{2}{c|}{SKCM} & \multicolumn{2}{c|}{Road Safety} & \multicolumn{2}{c|}{Compas} & \multicolumn{2}{c}{Upselling} \\
\cmidrule(){2-9}
& Mean & SD & Mean & SD & Mean & SD & Mean & SD \\
\midrule
RF & 0.763 & 0.017 & 0.796 & 0.002 & 0.778 & 0.003 & 0.791 & 0.007 \\
GB & 0.781 & 0.027 & 0.757 & 0.001 & 0.711 & 0.007 & 0.795 & 0.006 \\
MLP & 0.707 & 0.051 & 0.758 & 0.004 & 0.697 & 0.007 & 0.780 & 0.015 \\
LRF & 0.861 & 0.017 & 0.886 & 0.001 & 0.843 & 0.001 & 0.894 & 0.003 \\
\hline
CART & 0.718 & 0.033 & 0.721 & 0.003 & 0.680 & 0.007 & 0.744 & 0.010 \\
LR & 0.738 & 0.037 & 0.715 & 0.002 & 0.680 & 0.004 & 0.738 & 0.009 \\
LRT & 0.657 & 0.076 & 0.717 & 0.003 & 0.670 & 0.010 & 0.741 & 0.013 \\
\hline
SynthTree - INT & 0.814 & 0.032 & \textbf{0.763} & 0.013 & \textbf{0.743} & 0.008 & \textbf{0.891} & 0.011 \\
SynthTree - MLP & \textbf{0.835} & 0.028 & 0.760 & 0.010 & 0.742 & 0.007 & 0.842 & 0.016 \\
SynthTree - RF & 0.823 & 0.052 & 0.761 & 0.009 & 0.741 & 0.009 & 0.824 & 0.012 \\
SynthTree - GB & 0.814 & 0.043 & 0.757 & 0.007 & 0.734 & 0.009 & 0.839 & 0.023 \\
SynthTree - LRF & 0.827 & 0.032 & 0.752 & 0.004 & 0.740 & 0.009 & 0.818 & 0.001 \\
\hline
MLM-EPIC-INT & \textbf{0.740} & 0.051 & 0.778 & 0.002 & 0.752 & 0.001 & 0.818 & 0.012 \\
MLM-EPIC-MLP & 0.737 & 0.037 & \textbf{0.783} & 0.001 & \textbf{0.759} & 0.001 & \textbf{0.830} & 0.006 \\
MLM-EPIC-RF & 0.696 & 0.033 & 0.781 & 0.003 & 0.753 & 0.001 & 0.818 & 0.014 \\
MLM-EPIC-GB & 0.730 & 0.043 & 0.776 & 0.003 & 0.738 & 0.001 & 0.819 & 0.012 \\
\bottomrule
\end{tabular}
\caption{Compare classification performance by $AUC$ on four datasets. The average $AUC$'s and the standard deviations over five random runs (different splits into training and test sets) are reported. The best outcomes achieved respectively by SynthTree and MLM-EPIC methods are highlighted in bold.}
\label{tab:table1}
\end{table}

\begin{table}[!htb]
\centering
\begin{tabular}{l|cc|cc|cc|cc}
\toprule
\multirow{2}{*}{Model} & \multicolumn{2}{c|}{Cal Housing} & \multicolumn{2}{c|}{Bike Sharing} & \multicolumn{2}{c|}{Abalone} & \multicolumn{2}{c}{Servo}  \\
\cmidrule(){2-9}
& Mean & SD & Mean & SD & Mean & SD & Mean & SD \\
\midrule
RF & 0.514 & 0.009 & 43.506 & 1.583 & 2.157 & 0.040 & 0.365 & 0.157 \\
GB & 0.540 & 0.009 & 72.238 & 2.067 & 2.168 & 0.060 & 0.412 & 0.130 \\
MLP & 0.590 & 0.012 & 85.482 & 2.188 & 2.121 & 0.085 & 0.514 & 0.104 \\
LRF & 0.496 & 0.007 & 53.741 & 1.744 & 2.131 & 0.045 & 0.584 & 0.130 \\
\hline
CART & 0.725 & 0.016 & 60.624 & 2.664 & 2.964 & 0.107 & 0.545 & 0.151 \\
LR & 0.733 & 0.008 & 141.160 & 1.549 & 2.387 & 0.062 & 0.876 & 0.094 \\
LRT & 0.786 & 0.014 & 104.708 & 6.409 & 2.960 & 0.127 & 0.898 & 0.318 \\
\hline
SynthTree - INT & 0.716 & 0.034 & 54.730 & 2.176 & \textbf{2.137} & 0.048 & 0.533 & 0.105 \\
SynthTree - MLP & 1.566 & 0.889 & 74.256 & 3.867 & 2.141 & 0.066 & 0.816 & 0.141 \\
SynthTree - RF & 0.713 & 0.024 & \textbf{53.832} & 1.882 & 2.152 & 0.043 & \textbf{0.512} & 0.181 \\
SynthTree - GB & 0.713 & 0.027 & 65.547 & 1.467 & 2.158 & 0.039 & 0.582 & 0.097 \\
SynthTree - LRF & \textbf{0.711} & 0.030 & 79.978 & 7.782 & 2.149 & 0.054 & 0.686 & 0.105 \\
\hline
MLM-EPIC-INT & 0.639 & 0.033 & 68.522 & 1.445 & 2.174 & 0.055 & 1.054 & 0.113 \\
MLM-EPIC-MLP & 0.679 & 0.057 & 70.156 & 2.126 & 2.195 & 0.090 & 0.947 & 0.280 \\
MLM-EPIC-RF & 0.650 & 0.074 & 69.058 & 1.256 & \textbf{2.136} & 0.049 & \textbf{0.875} & 0.166 \\
MLM-EPIC-GB & \textbf{0.637} & 0.032 & \textbf{67.045} & 2.438 & 2.184 & 0.064 & 0.925 & 0.389 \\
\bottomrule
\end{tabular}
\caption{Compare regression performance by $RMSE$ on four datasets. The average $RMSE$'s and the standard deviations over five random runs (different splits into training and test sets) are reported. The best outcomes achieved respectively by SynthTree and MLM-EPIC methods are highlighted in bold.}
\label{tab:table2}
\end{table}

In Table~\ref{tab:table1}, we present the $AUC$ values for the four classification datasets. Table~\ref{tab:table2} details results for the four regression datasets. Our analysis reveals that black-box models generally outperform explainable models across these datasets. Notably, there is a significant variation in performance among the black-box models themselves. Specifically, LRF consistently registers the highest $AUC$ in the classification datasets, whereas RF records the best $RMSE$ in two of the four regression datasets. Both SynthTree and MLM-EPIC consistently outperform the explainable models in test data scenarios. Interestingly, in spite of SynthTree's focus on explainability, MLM-EPIC achieves better performance than SynthTree only for half of the datasets. In most scenarios, SynthTree and MLM-EPIC are outperformed by their respective co-supervision black-box models. 

\begin{figure}[!htb]
  \begin{minipage}{\textwidth}
    \centering
    \includegraphics[width=\textwidth]{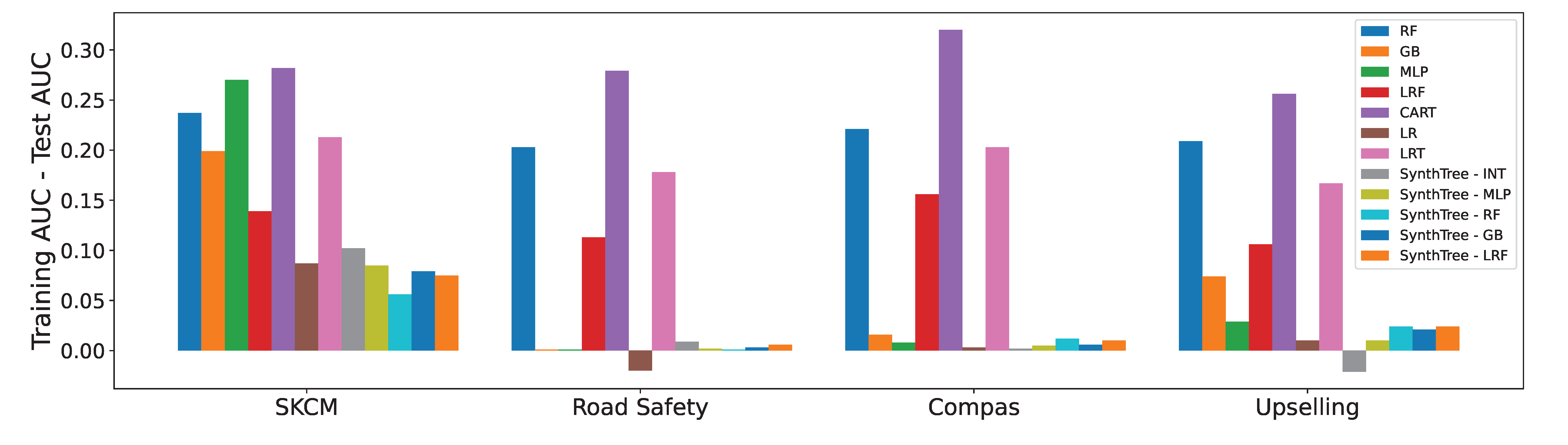}
    \caption*{(a) Classification}
  \end{minipage}
  \begin{minipage}{\linewidth}
    \centering
    \includegraphics[width=\linewidth]{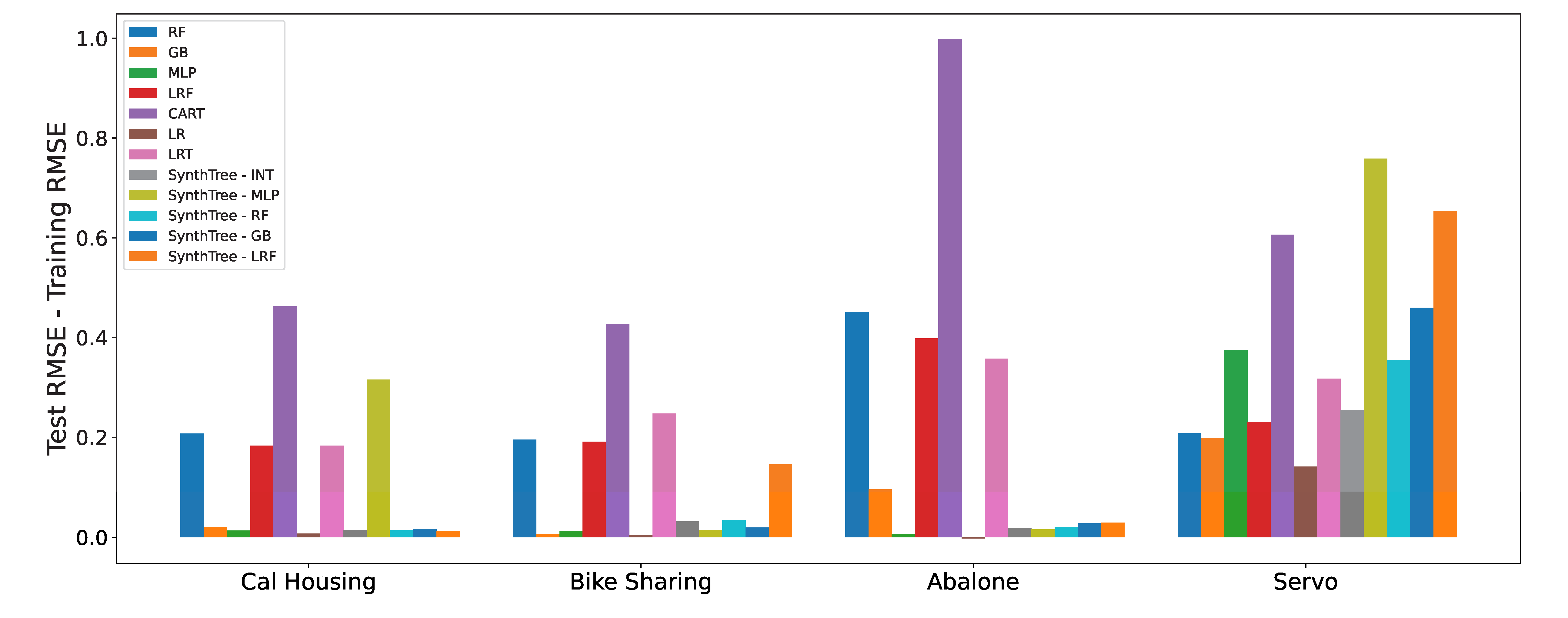}
    \caption*{(b) Regression}
  \end{minipage}
  \caption{Compare the level of overfitting in each model. Each bar represents one model, and each group of bars corresponds to a different dataset. (a) Classification: $AUC$ on the training set subtracted by that on the test set; (b) Regression: $RMSE$ on the test set subtracted by that on the training set. In both cases, a larger difference indicates more severe overfitting.}
      \label{fig:overfit}
\end{figure}

The strategy of employing multiple black-box models for co-supervision proves advantageous; it enhances SynthTree's performance, surpassing that of using a single black-box model in three classification and one regression datasets. 
For three of the four regression datasets, SynthTree-INT is ranked either first or second among the five SynthTree method variants. We also note that SynthTree-INT achieves competitive performance compared with the black-box models. For all four classification datasets and two of the four regression datasets, SynthTree-INT outperforms at least two of the four black-box models in test data evaluations. 
To examine resistance to overfitting, in Figure~\ref{fig:overfit}, we plot the difference between $AUC$'s (or $RMSE$'s) obtained on training and test datasets respectively. 
Figure~\ref{fig:overfit} demonstrates that compared with black-box models, CART, and LRT, SynthTree-INT often results in significantly lower overfitting for both classification and regression datasets. Although GB and MLP exhibit low overfitting for some datasets, they are not as consistent as SynthTree-INT in this regard. Furthermore, SynthTree-INT consistently yields lower overfitting compared to SynthTree when applied to individual black-box models.

\begin{table}[!htb]
\centering
\begin{tabular}{c|c|c|c}
\toprule
  Data & CART & LRT & SynthTree \\
  \hline
  SKCM & 51 & 126 & 4 \\
  \hline
  Bike Sharing & 13143 & 4899 & 128 \\
  \hline
  Compas & 3249 & 4787 & 16 \\
  \hline
  Abalone & 1990 & 1266 & 32 \\
  \hline
  Road Safety & 13713 & 34978 & 32 \\
  \hline
  Upselling & 443 & 1581 & 32 \\
  \hline
  Servo & 106 & 42 & 8 \\
  \hline
  Cal Housing & 15823 & 6581 & 8 \\
\bottomrule
\end{tabular}
\caption{The number of leaf nodes in trees generated by CART, LRT, and SynthTree respectively.}
\label{tab:treesize}
\end{table}


Tables~\ref{tab:table1} and~\ref{tab:table2} demonstrate that SynthTree consistently outperforms CART and LRT across various datasets. An intriguing aspect of this comparison lies in the explainability of the respective models. We thus examine the structural complexity of these tree-based models. Generally, a tree with fewer leaves is considered more interpretable. However, comparing CART to LRT or SynthTree is complicated due to their difference in models employed within leaf nodes: CART uses constant estimates, while LRT and SynthTree use linear models. Table~\ref{tab:treesize} displays the leaf node counts for trees generated by CART, LRT, and SynthTree. Notably, SynthTree consistently produces trees with significantly fewer leaves. While SynthTree's overall model complexity may not always be less than that of CART, especially when the tree size is dramatically reduced, it can be argued that a shallower tree with linear models at the leaf nodes is more interpretable than a deeper tree with constant leaf estimates.

\begin{figure}[!htb]
  \begin{minipage}[b]{0.42\textwidth}
    \centering
    \includegraphics[width=0.9\textwidth]{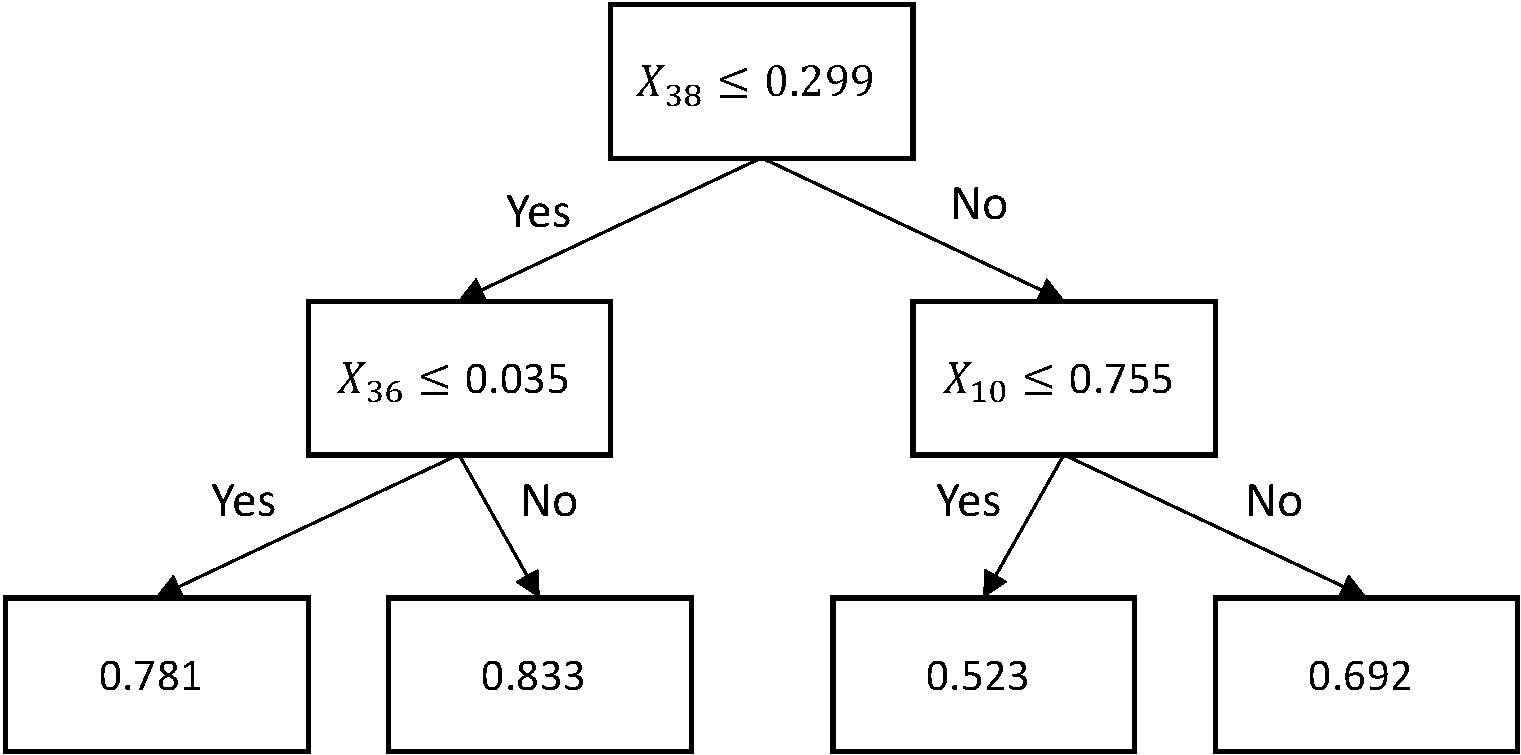}
    \vspace*{0.6in}
    \caption*{(a) SKCM}
  \end{minipage}%
  \begin{minipage}[b]{0.55\textwidth}
    \centering
    \includegraphics[width=0.9\textwidth]{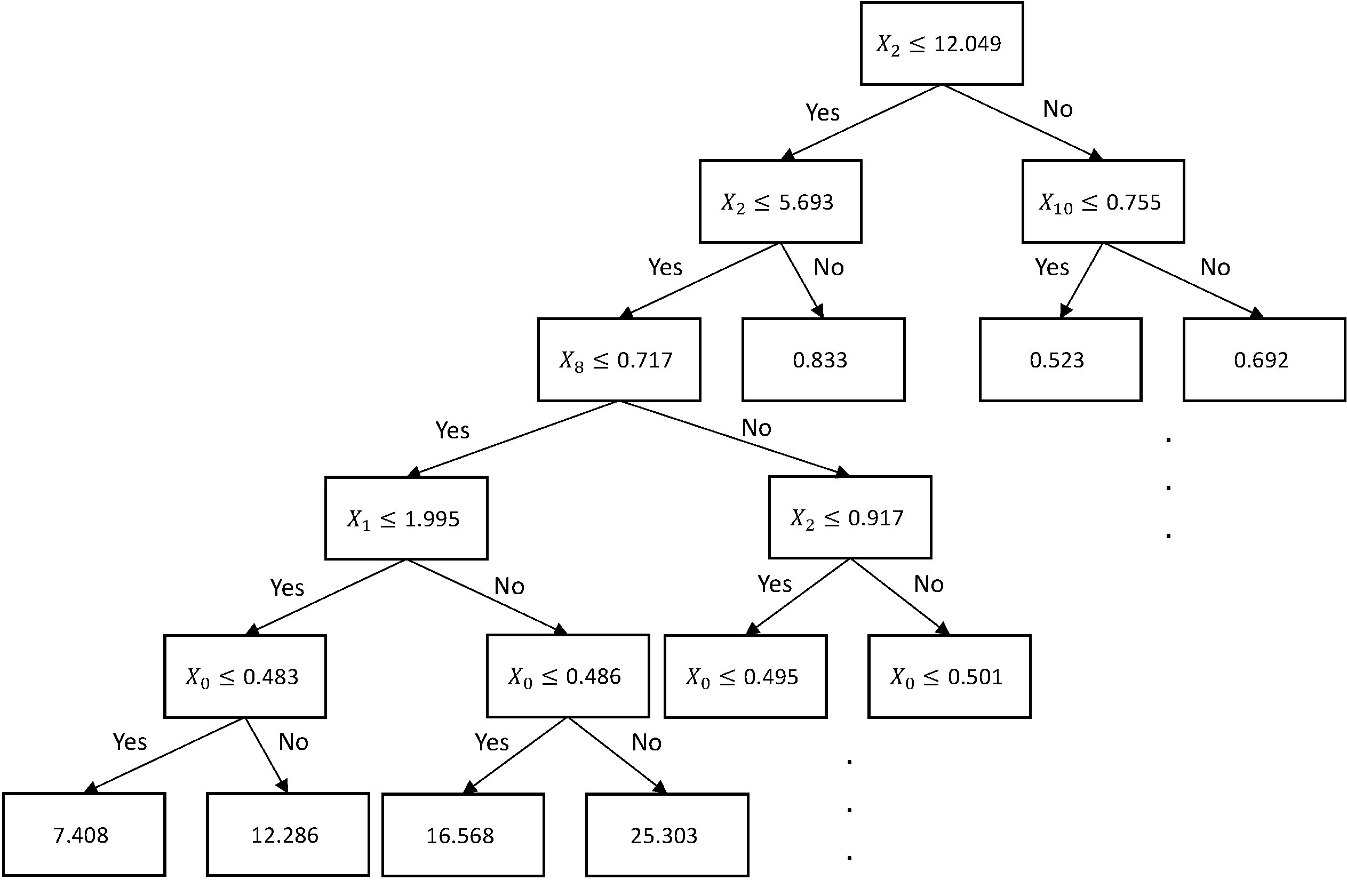}
    \caption*{(b) Bike Sharing}
  \end{minipage}
  \caption{The tree model obtained by SynthTree for (a) SKCM data, (b) Bike Sharing data. At each non-leaf node, we show the variable and the threshold used to make the split, while at each leaf node, we show the accuracy computed from the test data that fall in this node. As there are 48 leaf nodes in the SynthTree of Bike Sharing, the plot only shows a part of the tree for clarity.}
  \label{fig:skcm}
\end{figure}

Next, we demonstrate how to use the SynthTree model to elucidate the decision-making process. Figure~\ref{fig:skcm}(a) and (b) display respectively the tree model generated by SynthTree for the SKCM and Bike Sharing datasets. Take SKCM as an example. The feature space is segmented into four distinct regions (referred to as ELCs), defined by comparing two variables against their respective threshold values. Within each region, the interpretion task is straightforward since only a linear model is involved.

\begin{figure}[!htb]
  \begin{minipage}{\textwidth}
    \centering
    \includegraphics[width=\textwidth]{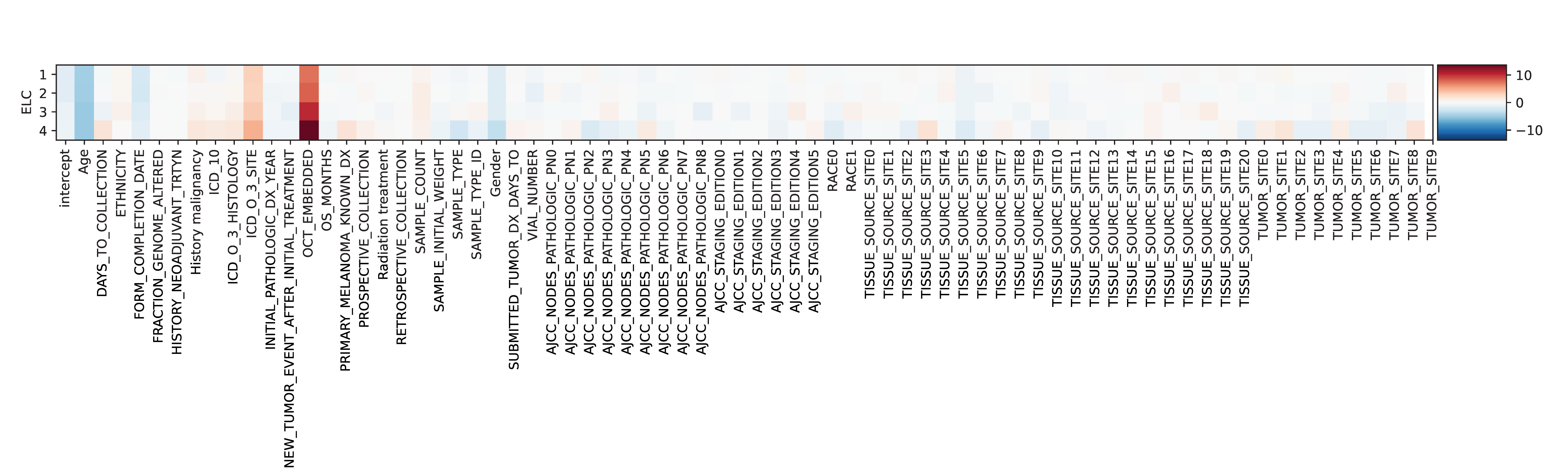}
    \caption*{(a) SKCM}
  \end{minipage}
  \begin{minipage}{\textwidth}
    \centering
    \includegraphics[width=\textwidth, height=0.3\textheight]{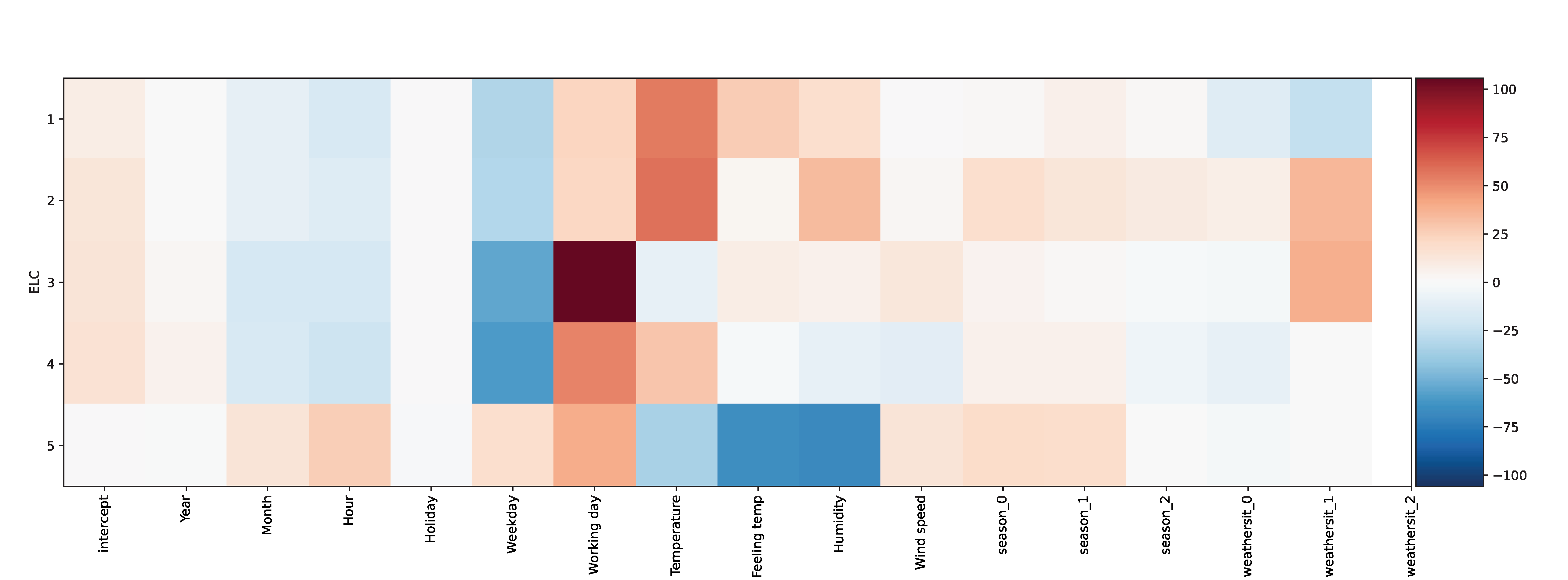}
    \caption*{(b) Bike Sharing}
  \end{minipage}
  \caption{Illustrate the regression coefficients in the linear model of each ELC using heat map. Every column in the heat map corresponds to one variable, and every row corresponds to one ELC. (a) SKCM; (b) Bike Sharing, in which case only                                     results for the five largest ELCs are shown.}
      \label{fig:skcm_1}
\end{figure}

\begin{figure}[!htb]
  \begin{minipage}{0.5\textwidth}
    \centering
    \includegraphics[width=\textwidth]{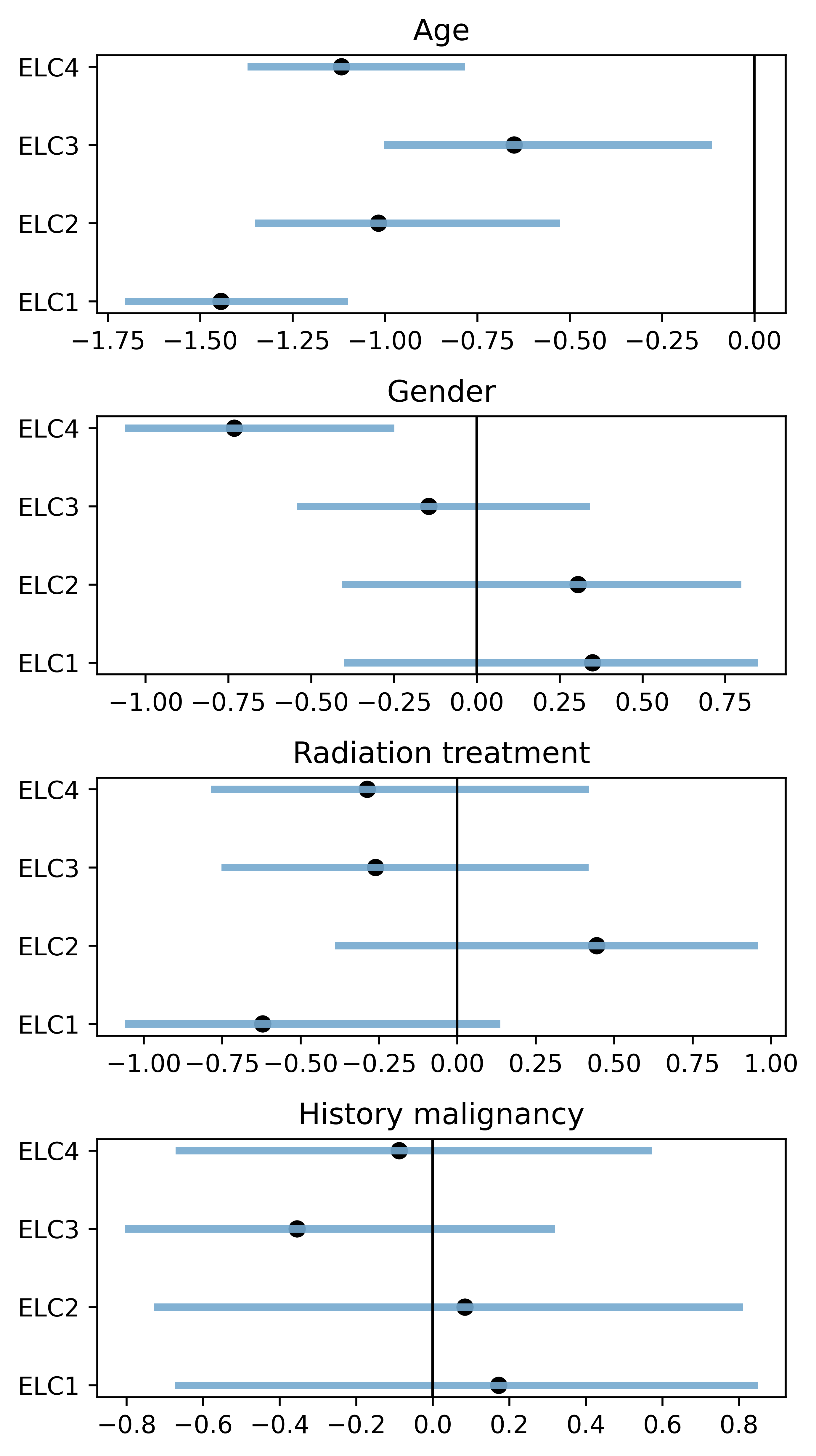}
    \caption*{(a) SKCM}
  \end{minipage}%
  \begin{minipage}{0.5\textwidth}
    \centering
    \includegraphics[width=\textwidth]{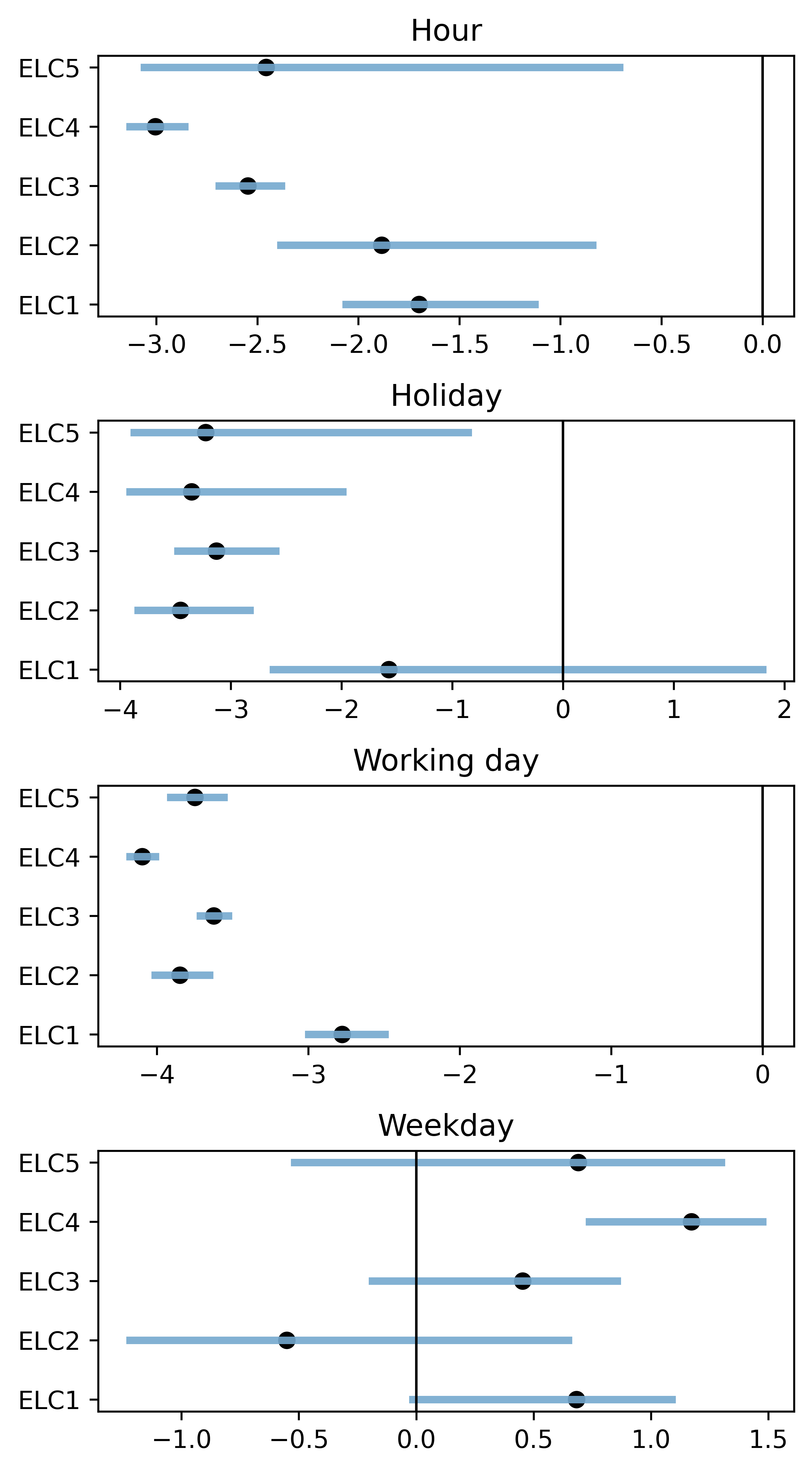}
    \caption*{(b) Bike Sharing}
  \end{minipage}
  \caption{Illustrate the linear models at the leaf nodes for SKCM (plots in the left column) and Bike Sharing data (plots in the right column) respectively. Every leaf node corresponds to one ELC. Regression coefficients and the confidence intervals are shown for four predictor variables in each dataset: Age, Gender, Radiation treatment, and History malignancy (from SKCM), and Hour, Holiday, Working Day, and Weekday (from Bike Sharing). In each plot, the coefficient and the confidence interval are shown separately for every ELC. }
      \label{fig:skcm_2}
\end{figure}

To showcase the linear models within each ELC, Figure~\ref{fig:skcm_1}(a) presents the linear coefficients for each predictor variable of SKCM. The figure consists of four rows of heat maps, each corresponding to a model within a specific ELC. In these heat maps, the saturation level of the color corresponds to the magnitude of a coefficient's absolute value. A more saturated color denotes a larger absolute value. Variables not included in the linear model are represented in white, highlighting their exclusion. Similarly, Figure~\ref{fig:skcm_1}(b) shows the linear models for the five largest ELCs of the Bike Sharing data.

Figure~\ref{fig:skcm_2}(a) and (b) display the linear coefficients and their confidence intervals for four variables of each dataset (SKCM and Bike Sharing). The plots reveal consistent patterns for ``AGE'' and ``SEX'', with their coefficients being negative across all four ELCs. This consistency indicates their similar roles in the models. In contrast, the coefficients for the other two variables exhibit sign change in different ELCs, signifying their divergent effects on the prediction.

In summary, Figures~\ref{fig:skcm},~\ref{fig:skcm_1}, and~\ref{fig:skcm_2} demonstrate how the SynthTree model facilitates interpretation of the SKCM and Bike Sharing data. This approach yields an interpretive complexity only slightly greater than that of a standard linear model. Furthermore, the variations observed in the models across different ELCs shed light on the data's heterogeneity, providing valuable insights.

\section{Conclusions and Future Work}
\label{sec:conclude}

In this study, we present SynthTree, an innovative algorithm developed to enhance the explainability of black-box prediction models while minimally impacting their accuracy. This approach combines the principles of decision trees and 
the training method of co-supervisory. Our experiments demonstrate that SynthTree can effectively leverage the benefits of both explainable statistical models and black-box models. In addition, we examined the strategy of using multiple black-box models for co-supervision and implemented this strategy on both SynthTree and MLM-EPIC. Considering the robust performance of MLM-EPIC-INT and SynthTree-INT, they can potentially be viewed as meta-learning approaches, rather than merely as tools for enhancing explainability. The results indicate that while these methods may not surpass all black-box models in performance, they often outperform a majority of the black-box models evaluated.
In practice, the optimal black-box model often varies depending on the dataset, and determining the best choice is not always straightforward. In this context, MLM-EPIC-INT and SynthTree-INT present an appealing alternative compared to the conventional approach of selecting the best-performing black-box model.


There are a few future directions to explore. 
For instance, it is interesting to assess systematically uncertainty in explanations derived from decision trees. In the realm of explainable machine learning, a prevailing yet unquestioned assumption is that a model's high predictive accuracy implies the reliability of its explanations. However, this assumption overlooks the possibility of multiple models attaining similar accuracy levels while significantly differing in their parameters. SynthTree offers a fresh perspective for exploring this issue. By directly comparing the ELCs and the linear models per ELC generated from various black-box models, we can probe questions such as whether distinct explanations represent different viewpoints or have arisen from nuances in estimation.
It is also worth exploring the applicability of SynthTree in scenarios where model-derived explanations can be directly evaluated by domain experts. Such applications are typically domain-specific and necessitate interdisciplinary collaboration. Engaging with experts across various fields could greatly enhance the practical utility and impact of SynthTree in real-world settings.

\bibliographystyle{apalike}
\bibliography{ref.bib}

\end{document}